\documentstyle[psfig,prd,aps]{revtex}
\begin{document}
\draft

\title{Scale invariance and critical gravitational collapse}

\author{David Garfinkle\thanks {Email: garfinkl@vela.acs.oakland.edu} 
and Ken Meyer}
\address{
\centerline{Department of Physics, Oakland University,
Rochester, Michigan 48309}}

\maketitle

\null\vspace{-1.75mm}

\begin{abstract}

We examine ways to write the Choptuik critical solution as the
evolution of scale invariant variables.  It is shown that a system
of scale invariant variables proposed by one of the authors\cite{me1}
does not evolve periodically in the Choptuik critical solution.  We find a
different system, based on maximal slicing.  This system does evolve
periodically, and may generalize to the case of axisymmetry or of no
symmetry at all.
 
\end{abstract}
%\twocolumn
 
 
\section{Introduction}

Scaling behavior, as first found by Choptuik\cite{choptuik}, occurs at 
and near the
threshold of black hole formation in the gravitational collapse of many 
types of
matter\cite{evans1,eardley,garfinkle,evans2,stewart,bizon,traschen,piran3}.  
For a
one parameter family of initial data slightly above the threshold, the 
mass of the
black hole scales like $ {{(p - p*)}^\gamma }$.  Here
$p$ is the parameter, $p*$ is its critical value and $\gamma $ is 
a constant that
depends on the type of matter, but not on the family of data.  For initial
data slightly below the threshold, the maximum curvature scales like
${{({p*} - p)}^{- 2 \gamma }}$ where $\gamma $ is the same constant as in the
black hole mass scaling law.\cite{me2}  For some types of matter, the
critical solution ($p=p*$) has periodic self-similarity: after a certain
amount of time, the metric and matter variables repeat themselves with the
scale of space shrunk.  For other types of matter, the critical solution
has exact self-similarity: only the overall spatial scale changes with time.

Most of the work done on scaling has assumed spherical symmetry.  However,
Abrahams and Evans have found scaling in the collapse of axisymmetric
gravity waves\cite{evans2}.  In addition, perturbative work of Gundlach on
fluid collapse assumes no symmetries at all.\cite{gundlach4}  
Thus (for certain types of matter) scaling seems to be a generic property 
of critical gravitational collapse. 

The scaling of black hole mass (and of maximum curvature) has been
explained\cite{hara,gundlach1,piran2} subject to the following 
assumptions: (i) the
critical solution is periodically self-similar or exactly 
self-similar and (ii) the
critical solution has exactly one unstable mode.  What remains to be
explained is why the critical solution is periodically self-similar (or
exactly self-similar).  It has been noted\cite{me1,hara,gundlach1} 
that the property
of periodic self-similarity bears a striking resemblance to periodicity of a
limit cycle of a dynamical system.  Correspondingly, exact self-similarity
of a critical solution resembles a limit point of a dynamical system. 
Furthermore, one can turn ``resembles'' to ``is'' by writing the Einstein
field equations as a dynamical system of scale invariant variables.

To write Einstein's equations as a dynamical system requires a choice of
foliation (time slicing) for the spacetime.  Some foliations will be
compatible with the periodic self-similarity and some will not.  More
precisely, periodic self-similarity means that there is a diffeomorphism
$\zeta $ and a number $\Delta $ such that $ {\zeta ^*} \left ( {g_{ab}}
\right ) = {e^{ - 2 \Delta }} \, {g_{ab}} $.  A foliation is a
one parameter family of spacelike (or null) hypersurfaces $\Sigma (t)$. 
For a periodically self-similar spacetime, call a foliation compatible
provided that for each $t_0$ there is a
$t_1$ such that 
$\zeta \left ( \Sigma (t_0) \right ) = \Sigma ({t_1})$.  In more physical
terms a foliation is compatible if for
each slice, there is a later slice that is identical except for an overall
change in the scale of space.  For a given periodically self-similar
spacetime, there are many foliations that are compatible.  Simply choose any
initial hypersurface $\Sigma ({t_0})$.  Carry it forward by the
diffeomorphism $\zeta $ to form the hypersurface 
$ \Sigma ({t_1} )$.  For $ {t_0} < t < {t_1} $, interpolate the
hypersurfaces in any smooth way.  Then extend that family to the whole
spacetime using $\zeta $ and $\zeta ^{- 1}$.  There are also many 
foliations that are not compatible.  
Thus compatibility is a property of the entire foliation, or
equivalently of the prescription for the lapse.  Given any initial
slice, there is some compatible foliation that contains that slice.  A
dynamical systems explanation of Choptuik scaling requires a
compatible foliation.

Note that compatibility is a condition on the lapse and is independent
of the choice of shift.  However, one may want to choose the
evolution vector field so that the diffeomorphism
$\zeta$ has the form $t\to t+ \Delta$ with the spatial coordinates
unchanged.  This requires both a compatible foliation and a
condition on the shift.

Most of the work done on Choptuik scaling has assumed spherical symmetry. 
Here the metric can be put in the form\cite{choptuik}
$$
d {s^2} = - \, {\alpha ^2} \, d {t^2} \; + \; {a^2} \, d {r^2} \; + \;
{r^2} \, d {\Omega ^2} \; \; \; .
\eqno(1)
$$
Here $\alpha $ and $a$ are functions of $r$ and $t$; $d {\Omega ^2}$ is the
unit two-sphere metric and $r$ is the usual area coordinate.  The surfaces
of constant $t$ are orthogonal to the surfaces of constant $r$.  Usually
the condition $ \alpha = 1 $ at $r = 0 $ is imposed so $t$ is proper time
at the position of the central observer.  This foliation is
compatible.  

A spherically symmetric metric can also be put in the form\cite{christodoulou}
$$
d {s^2} = - \, g \, d u \, ( {\bar g} \, d u \, + \, 2 \, d r ) \; + \; 
{r^2} \, d {\Omega ^2} \; \; \; .
\eqno(2)
$$
Here $r$ and $d {\Omega ^2} $ are as in equation (1) and $g$ and $\bar g$
are functions of $r$ and $u$.  The ``time'' coordinate $u$ is
constant along outgoing light rays and equal to proper time at the
position of the central observer.  This foliation is compatible.

A metric or matter variable is scale invariant provided that it is unchanged
when ${g_{ab}} \to k \, {g_{ab}} $ for any positive constant $k$.  The
quantities $\alpha $ and $a$ in equation (1) are scale invariant: under a
scale transformation, $\alpha $ and $a$ have the same form with only the
overall scale of $r$ and $t$ changed.  Similarly, the quantities $g$ and
$\bar g$ in equation (2) are scale invariant.  In spherical symmetry, the
metric is determined by the matter.  The type of matter treated by
Choptuik\cite{choptuik} is a massless, minimally coupled scalar field
$\phi$.  Initial data for this system consists of the field and its normal
derivative at a moment of time.  The matter variables used in\cite{choptuik}
are $(r/a) \, \partial \phi /\partial r$ and $(r/\alpha ) \, \partial \phi
/ \partial t$.  These variables are scale invariant.  For the same physical
system, in the coordinates of equation (2) the usual matter variable is $h
\equiv \partial (r \phi )/\partial r$, which is scale invariant.  Scale
invariant variables have also been found for other types of
matter.\cite{gundlach2,gundlach3}

While compatible foliations are known
in the spherically symmetric case, Choptuik scaling is a generic phenomenon
that does not seem to depend on symmetry.  Therefore, one would like to find
a compatible foliation for the general case.  The foliations of equations (1)
and (2) depend, in an essential way, on spherical symmetry.  The constant
$t$ slices of equation (1) are defined as orthogonal to the spherical area
coordinate $r$.  The constant $u$ slices of equation (2) are the null
cones of the observer left invariant by the spherical symmetry.  Thus,
there does not seem to be any natural way to generalize these foliations to
the case of axisymmetry or of no symmetry.

Thus, what should be done is to use a slicing condition that makes no
reference to symmetry, find a set of scale invariant variables associated
with that foliation, and check that the foliation is compatible.  In practice,
one can check compatibility on a case by case basis.  That is, given a
particular critical solution, say as the output of a numerical code, one
can check numerically whether a particular foliation is compatible.
In this paper, we will explicitly check compatibility only with the
Choptuik critical solution: that is, the case of a spherically symmetric,
minimally coupled scalar field.

In reference\cite{me1} a slicing condition was proposed that makes no
reference to symmetry, and a set of scale invariant metric and matter
variables associated with this foliation was found.  However, it was not
known whether this foliation was compatible.  This paper
provides an answer to that question.  In section 2 we present the numerical
methods used in this study and consider the foliation of
reference\cite{me1}. We show that this foliation is not compatible with the
Choptuik critical solution.  In section 3 we find a different system of scale
invariant variables based on maximal slicing, and show that maximal slicing
is compatible with the Choptuik critical solution.  We then consider the
question of whether maximal slicing is compatible in the case of
axisymmetry or no symmetry.

\section{Numerical Methods}

Typically, in numerical relativity, one thinks of the slicing condition as
a part of the numerical code that evolves the metric variables to find the
spacetime.  Thus, the prospect of analyzing the Choptuik critical solution
using several different slicing conditions seems daunting.  
A numerical treatment of the
Choptuik critical solution requires great accuracy and stability and a
large range of spatial scales.  It would be difficult to write a code that
does these things well in an arbitrary slicing.  Fortunately, this is not
necessary.  Given the data from a code that uses one slicing condition, one
can produce numerically the foliations of other slicing conditions.  

We begin
with numerical data for the metric and scalar field for the Choptuik
critical solution in the coordinates of equation (2), as found in
reference\cite{garfinkle} (using methods based on the work of
Christodoulou\cite{christodoulou} and of Goldwirth and Piran\cite{piran1}).  
A spherically symmetric slice in these  coordinates is given by
$u=u(r)$.  
To produce a slice in these coordinates, we must express the slicing
condition as a differential equation for $u(r)$ and then numerically
solve that equation.
For the $t=const.$ slices of
equation (1), ${\nabla _a} t$ is orthogonal to ${\nabla _a} r$.
Therefore these slices are given by
$$
{{d u} \over {d r}} = - \; {1 \over {\bar g}} \; \; \; .
\eqno(3)
$$
To solve this equation numerically, for each $u$ we begin at $r=0$
and integrate outward using a Runge-Kutta method.  For each $(r,u)$ value we
find the value of $ \bar g $ by interpolation using the
values that we have from the output of the numerical code of
reference\cite{garfinkle}.  

Now suppose that we wish to find the maximal slices.  For a surface
$u=u(r)$, the extrinsic curvature is 
$$
K = {W \over {2 g}} \; {{\partial g} \over {\partial u}} \; + \; 
{g \over 2} \; {W^3} \; {{\partial {\bar g}}
\over {\partial u}} \; + \; \left ( {1 \over {g W}} \; - \; {\bar g} \, W
\right ) \; \left ( {1 \over {4 g}} \; {{\partial g} \over {\partial r}} \;
+ \; {1 \over r} \right ) 
$$
$$
 - \; {1 \over 4} \; \left ( g \, {\bar g} \,
{W^3} \, + \, 3 \, W \right ) \; {{\partial {\bar g}} \over {\partial r}} \;
- \; {{g \, {W^3}} \over {{(du/dr)}^3}} \; {{{d^2} u} \over {d {r^2}}} 
\eqno(4)
$$
Where $W$ is given by
$$
W \equiv {{\left [ - \, g \, \left ( {\bar g} \, + \, 2 {{[du/dr]}^{-1}}
\right ) \right ] }^{-1/2}} \; \; \; .
\eqno(5)
$$
Thus the equation $K=0$ becomes
$$
{{{d^2} u} \over {d {r^2}}} = {1 \over g} \; {{\left ( {1 \over W} \; {{du}
\over {d r}} \right ) }^3} \; \bigg [  {W \over {2 g}}  \; {{\partial g} \over
{\partial u}} \; + \; {g \over 2} \; {W^3} \; {{\partial {\bar g}}
\over {\partial u}} \; + \; \left ( {1 \over {g W}} \; - \; {\bar g} \, W
\right ) \; \left ( {1 \over {4 g}} \; {{\partial g} \over {\partial r}} \;
+ \; {1 \over r} \right )
$$
$$
 - \; {1 \over 4} \; \left ( g \, {\bar g} \,
{W^3} \, + \, 3 \, W \right ) \; {{\partial {\bar g}} \over {\partial r}}
\bigg ] \; \; \; .
\eqno(6)
$$
For each $u$ we begin at $r=0$ and produce each surface by a Runge-Kutta
integration outward using equation (6).  Again, for each $(r,u)$ value, the
quantities $ g, {\bar g}$ and their derivatives are interpolated from 
the gridpoint
values of these quantities.

Now suppose that our slicing condition involves a dynamical evolution of
the lapse.  Given the lapse $N$, we evolve the slice, and given the
conditions on the slice, we evolve the lapse.  For a slice, $u = u(r)$, the
unit normal vector is 
$$
{n^a} = W \; {{\left ( {\partial \over {\partial u}}\right ) }^a} \; - \;  W
\; \left [ {\bar g} \; + \; {1 \over {du/dr}}\right ] \; {{\left ( {\partial
\over {\partial r}}\right ) }^a} \; \; \; .
\eqno(7)
$$
Since we wish only to know how the slices evolve, we can without loss of
generality choose the shift to vanish.  The coordinates of the slice then
evolve by
$$
{\dot u} = N \, W \; \; \; ,
\eqno(8a)
$$
$$
{\dot r} = - \, N W \; \left [ {\bar g} \; + \; {1 \over {du/dr}}\right ] \;
\; \; .
\eqno(8b)
$$
Here an overdot denotes derivative with respect to the evolution vector
field $N \, {n^a} $.  

For the system of reference\cite{me1}  The lapse
evolves by
$$
{\dot N} = {1 \over 3} \; {N^2} \, K \; \; \; .
\eqno(9)
$$
Let $({h_{ab}}, \, {K_{ab}}, \, {D_a})$ be respectively the intrinsic
metric, extrinsic curvature and derivative operator of a slice. The scale
invariant metric variables associated with this foliation are 
${{\tilde h}_{ab}} \equiv {N^{-2}} \, {h_{ab}}, \, {{\tilde K}_{ab}} \equiv
{N^{-1}} \, \left ( {K_{ab}} \, - \, [K/3] \, {h_{ab}} \right )$ and 
${\omega _a}
\equiv {N^{-1}} \, {D_a} N $.

We choose as the initial slice, a $t = const. $ slice generated using
equation (3).  The initial lapse is chosen to be $1$.  The slice and lapse
are then evolved using equations (8a), (8b) and (9) 
to produce a foliation of the spacetime.

We would now like to know whether this foliation is compatible.  In a
compatible foliation, the scale invariant variables repeat their behavior. 
Thus at corresponding slices, the scalar field $\phi $ as a function of 
$ r / {\sqrt N} $ should be the same.  To identify which slices are
``corresponding,'' define $\phi _0$ to be the value of $\phi $ at $r=0$.
Then $\phi _0$ repeats itself and the slices that contain repeating values
of $\phi _0$ are corresponding slices.  

\begin{figure}[bth]
%fig1
\begin{center}
\makebox[4in]{\psfig{file=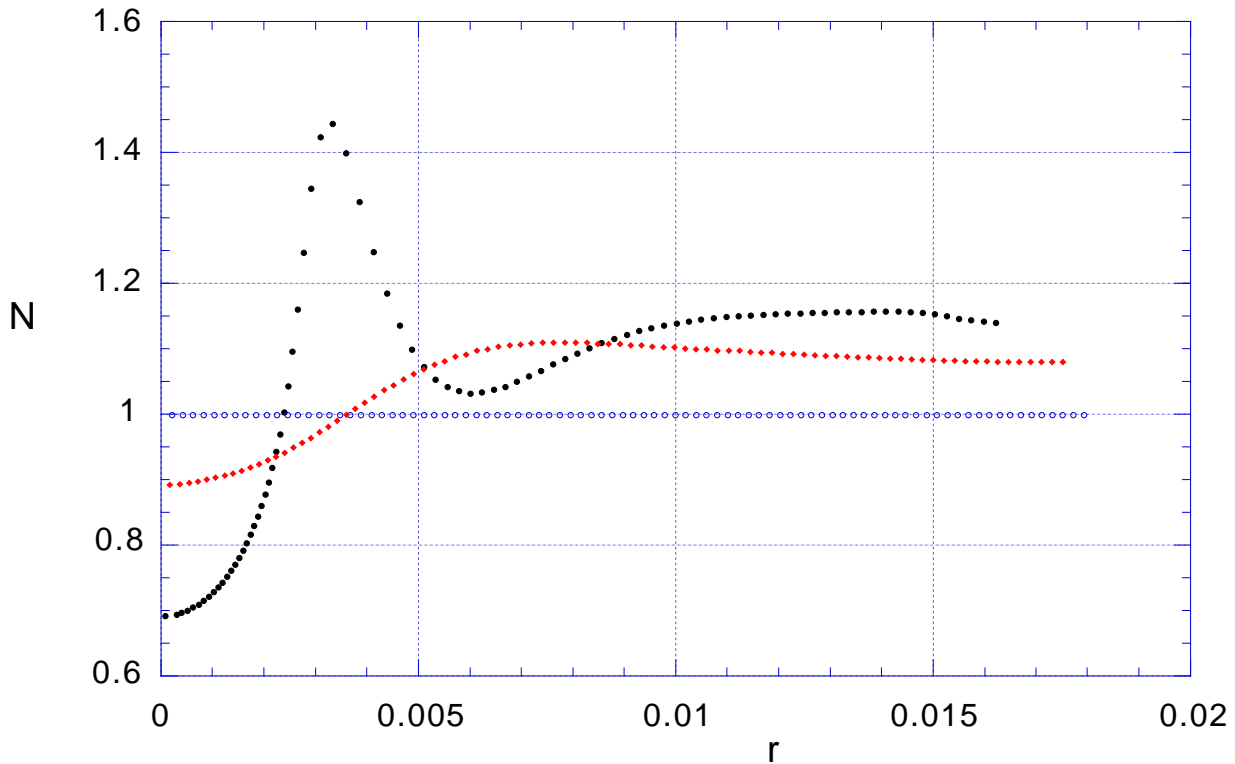,width=6.5in}}
\caption{The lapse is plotted at the initial time and two later times.
Note the steep peak in the lapse at the latest of these times.}
\label{fig1.eps}
\end{center}
\end{figure}

Note from equation (9) that positive $K$ leads to an increasing lapse 
and, in fact,
can cause the lapse to blow up in finite time.  Figure 1 shows the lapse 
on three
different slices.  Note the steep peak in the lapse at the latest of 
these times. 
Shortly after this time, the lapse blows up.  In the sequence of 
slices from the
begining to the blow up of the lapse, $\phi _0$ does not even go 
through one cycle.
Thus this foliation does not even exist long enough for a comparison 
of corresponding
slices to take place.  Therefore this foliation is not compatible.

\section{Maximal Slicing}

A dynamical systems explanation of Choptuik scaling requires a compatible
foliation and a set of scale invariant variables.  For the Einstein-scalar
equations, initial data on a slice $\Sigma $ are $( {h_{ab}}, \, {K_{ab}},
\phi , \, P )$ where $h_{ab}$ is the intrinsic metric, $K_{ab}$ is the
extrinsic curvature, $\phi $ is the scalar field and $P$ is its normal
derivative.
(recall that for $n_a$ the unit one-form normal to $\Sigma$ we have
${h_{ab}}={g_{ab}}+{n_a}{n_b}, \; {K_{ab}}={{h_a}^c}{\nabla_c}{n_b}$
and $P={n^a}{\nabla_a}\phi$).   
If $({g_{ab}},\, \phi )$ is a solution of the Einstein-scalar
equations and $c$ is any positive constant, then $ ( {c^2} {g_{ab}}, \,
\phi )$ is also a solution. 
In this new solution, the unit normal one-form is now $c {n_a}$.  It
then follows that the initial data for this solution are
$ ( {c^2} {h_{ab}}, \, c {K_{ab}}, \, \phi , \, P/c )$.  Since a maximal
slice in $g_{ab}$ is also a maximal slice in $ {c^2} {g_{ab}}$, it is
natural to define the following equivalence relation among maximal initial
data sets:  a set $ ({{\tilde h}_{ab}}, \, {{\tilde K}_{ab}} , \, {\tilde
\phi }, \, {\tilde P}) $ is equivalent to a set $( {h_{ab}}, \, {K_{ab}},
\phi , \, P )$ provided that there is a positive constant $c$ such that
$$
({{\tilde h}_{ab}}, \, {{\tilde K}_{ab}} , \, {\tilde \phi }, \, {\tilde
P}) = ( {c^2} {h_{ab}}, \, c {K_{ab}}, \, \phi , \, P/c ) \; \; \; .
\eqno(10)
$$

Our set of scale invariant variables is an equivalence class of initial
data.  Is this set complete?  That is, does an equivalence class of initial
data contain enough information to determine its evolution?  The answer is
yes: simply take a member of the equivalence class, evolve it, and on each
slice take its equivalence class.  It does not matter which member of the
equivalence class we pick, since in each case the metric that we produce
differs only by an overall constant, leading to the same maximal slices and
the same equivalence classes.  

For the case of vacuum spacetimes, the scale invariant variables are simply
equivalence classes of $({h_{ab}}, \, {K_{ab}})$.  For types of matter other
than the scalar field, one must add scale invariant variables corresponding
to initial data for the matter fields.

An equivalence class is a somewhat abstract variable.  Is there a way to
make these scale invariant variables more concrete?  In some cases, the
answer is yes.  Suppose that $ {K^{ab}} {K_{ab}} $ is bounded on a slice 
$\Sigma $ (as will be the case if $\Sigma $ is asymptotically flat) and
define $\lambda \equiv {\sup _\Sigma } {{\left ( {K^{ab}} {K_{ab}} \right )
}^{1/2}}$.  Then under the transformation ${g_{ab}} \to {c^2} \, {g_{ab}} $
we have $\lambda \to \lambda /c$.  Now define
$$
( {{\bar h}_{ab}}, \, {{\bar K}_{ab}}, \, {\bar \phi}, \, {\bar P}) 
\equiv 
({\lambda ^2} \, {h_{ab}}, \, \lambda \, {K_{ab}}, \, \phi , \, P/\lambda
) \; \; \; .
\eqno(11)
$$
Then under ${g_{ab}} \to {c^2} \, {g_{ab}} $ the set $( {{\bar h}_{ab}}, \,
{{\bar K}_{ab}}, \, {\bar \phi}, \, {\bar P}) $ remains unchanged.  In
fact, this set is isomorphic to equivalence classes of $({h_{ab}}, \,
{K_{ab}}, \, \phi , \, P)$.  Thus for concreteness, in the case where 
${K^{ab}} \, {K_{ab}}$ is bounded on each slice, we can use 
$( {{\bar h}_{ab}}, \,
{{\bar K}_{ab}}, \, {\bar \phi}, \, {\bar P}) $ as our set of scale
invariant variables.

\begin{figure}[bth]
%fig1
\begin{center}
\makebox[4in]{\psfig{file=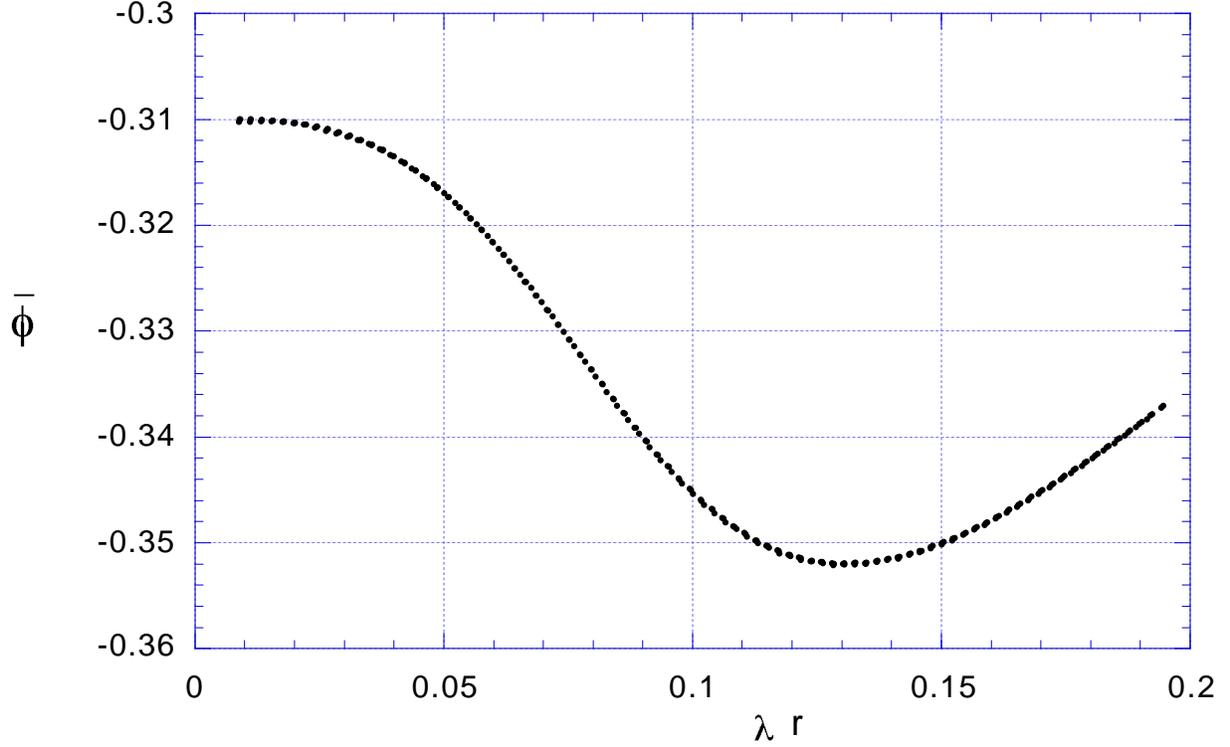,width=6.5in}}
\caption{The scalar field is plotted as a function of rescaled radius}
\label{fig2.eps}
\end{center}
\end{figure}

\begin{figure}[bth]
%fig1
\begin{center}
\makebox[4in]{\psfig{file=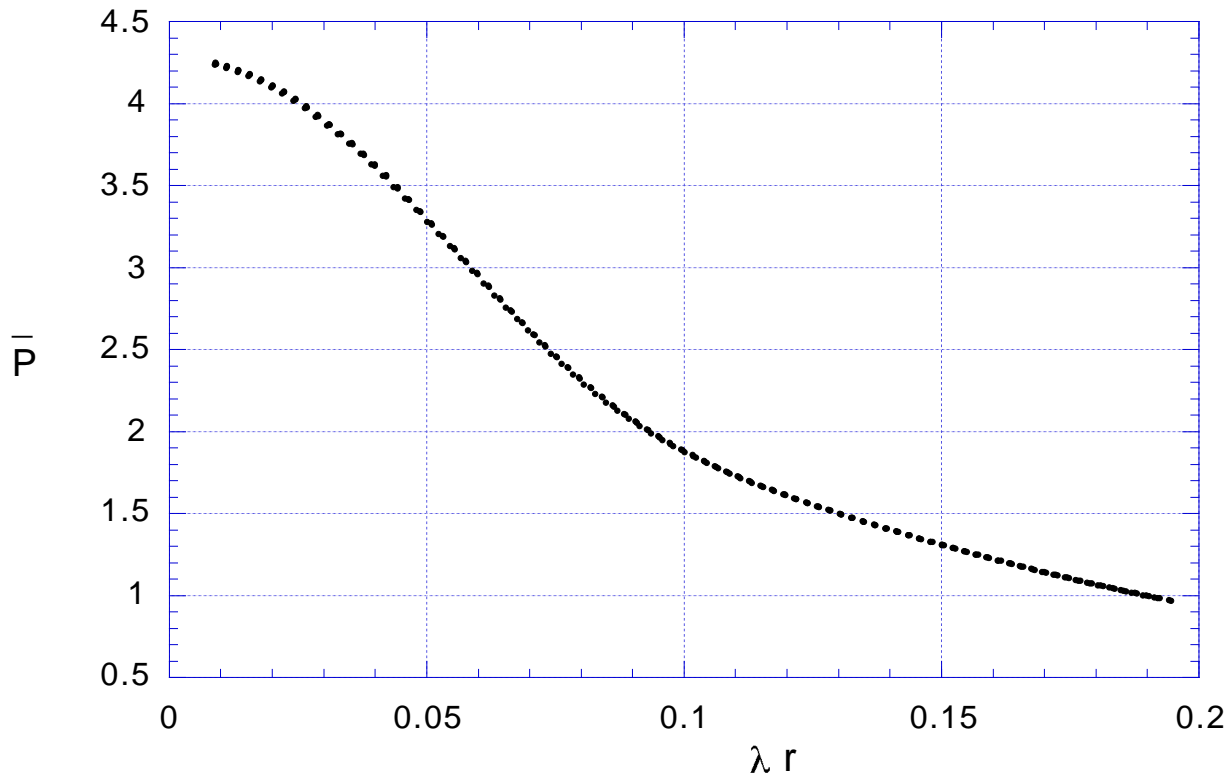,width=6.5in}}
\caption{The rescaled time derivative of the scalar field is plotted as
a function of rescaled radius}
\label{fig3.eps}
\end{center}
\end{figure}

We now consider whether maximal slicing is compatible with the Choptuik
critical solution.  The maximal slices are generated numerically as
described in the previous section.  Compatibility of the foliation means that
on corresponding slices, $\bar \phi $ and $\bar P$ as functions of $\lambda
r$ are the same.  Note that this is all that one needs to check, since in
spherical symmetry the metric variables are determined by the matter
variables.  In figure 2, $\bar \phi $ is plotted vs. $\lambda r$ for
three corresponding slices (those with $\phi _0$ halfway between zero and
its minimum).  These slices correspond to 
$\lambda = 22.475, \, 706.33$ and $22361$.  Note that the three different 
curves
are essentially indistinguishable, so $\bar \phi $ is the same on 
corresponding
slices.  In figure 3, $\bar P$ is plotted for these slices.  
Once again, the three
curves are essentially indistinguishable, so $\bar P$ is the same on all
three slices.  It then follows that maximal slicing is compatible with
the Choptuik critical solution.

This numerical study looks only at the case of the massless, minimally
coupled, spherically symmetric scalar field.  Nonetheless, we now argue
that maximal slicing is always compatible in the case of spherical
symmetry, and may be compatible in the case of axisymmetry or no symmetry.
Let $g_{ab}$ be a periodically self-similar metric with associated
diffeomorphism $\zeta $.  Let $\Sigma $ be a maximal slice.  Since 
$ {\zeta ^*} ( {g_{ab}} ) = {e^{ - 2 \Delta }} \, {g_{ab}} $, it follows
that $\zeta ( \Sigma ) $ is a maximal slice.  Therefore there must be a
compatible maximal slicing that includes $\Sigma $.   If the metric is
spherically symmetric, there is a unique spherically symmetric maximal 
slicing.  Therefore, this maximal slicing must be compatible.  In the case
of axisymmetry or no symmetry, things are more complicated because maximal
slicing is no longer unique.  The condition on the lapse $N$ that preserves
maximal slicing is
$$
{D_a} {D^a} N \; + \; N \, \left [ {R_{ab}} {n^a} {n^b} \, + \, R \, 
- \, {^{(3)}R} 
\right ] = 0 \; \; \; .
\eqno(12)
$$
Solutions of this equation are determined by boundary conditions.  For
gravitational collapse, the usual condition is that $N \to 1 $ at infinity.
Perhaps this condition leads to a compatible maximal slicing.  The work of
reference\cite{evans2} on scaling in axisymmetric vacuum collapse uses maximal
slicing with this condition on the lapse and finds that this foliation 
is compatible.
It therefore seems likely that maximal slicing with $N \to 1$ at infinity in
general leads to a compatible slicing.  In any case, there is some 
maximal slicing
that is compatible, even in the absence of symmetry.  In this foliation, 
with our scale
invariant variables, critical gravitational collapse is simply a 
limit cycle of a
dynamical system.

\section{Acknowledgements}

We would like to thank Alan Rendall, Carsten Gundlach and 
Chuck Evans for helpful
discussions. This work was partially supported by NSF grant PHY-9722039 
and by a
Cottrell  College Science Award of Research Corporation to Oakland University.  
Part of this work was performed by KM in partial fulfillment 
of the requirements
for the MS in Physics degree at Oakland University.

\vfill\eject 
\end{document}